\documentclass[twocolumn,aps,pra,showpacs,bibfootnote,preprintnumbers, mathtools, superscriptaddress, amsmath,amssymb]{revtex4-2}
\usepackage{graphicx}
\usepackage{optidef}
\usepackage{xcolor}
\usepackage{hyperref}
\usepackage{caption}
\usepackage{subcaption}
\usepackage{changes}

\begin{document}

\title{Superoscillations Made Super Simple}

\author{Gerard McCaul}
\email{gmccaul@tulane.edu}
\affiliation{Tulane University, New Orleans, LA 70118, USA}

\author{Peisong Peng}
\email{ppeng@tulane.edu}
\affiliation{Tulane University, New Orleans, LA 70118, USA}

\author{Monica Ortiz Martinez}
\email{mortizmartinez@tulane.edu}
\affiliation{Tulane University, New Orleans, LA 70118, USA}

\author{Dustin Lindberg
\href{https://orcid.org/0001-5335-7941}{\includegraphics[scale=0.05]{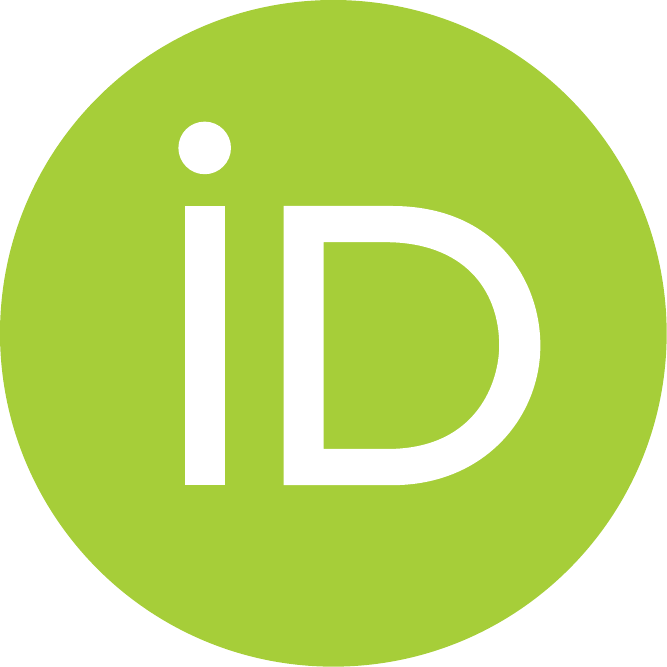}}}
\email{dlindberg@tulane.edu}
\affiliation{Tulane University, New Orleans, LA 70118, USA}

\author{Diyar Talbayev}
\email{dtalbaye@tulane.edu}
\affiliation{Tulane University, New Orleans, LA 70118, USA}

\author{Denys I. Bondar
\href{https://orcid.org/0000-0002-3626-4804}{\includegraphics[scale=0.05]{orcidid.pdf}}}
\email{dbondar@tulane.edu}
\affiliation{Tulane University, New Orleans, LA 70118, USA}






\begin{abstract}
In ordinary circumstances the highest frequency present in a wave is the highest frequency in its Fourier decomposition. It is however possible for there to be a spatial or temporal region of the wave which locally oscillates at a still greater frequency, in a phenomenon known as superoscillation \cite{Berry_2019}. Superoscillations find application in wide range of disciplines  \cite{Eliezer:14, Eberly_2016,zarkovsky_transmission_2020}, but at present their generation is based upon constructive approaches which are difficult to implement. Here we address this, exploiting the fact that superoscillations are a product of destructive interference to produce a prescription for generating superoscillations from the superposition of arbitrary waveforms. As a first test of the technique, we use it to combine four THz laser fields generated by periodically poled Lithium Niobate. From this, we are able to predict and observe for the first time THz optical superoscillations in the temporal domain. The ability to generate superoscillations in this manner has potential application in a wide range of fields. It may for example contribute to the experimental realization of the complex pulses required by quantum control, and the generation of attosecond pulses without resorting to nonlinear processes.
\end{abstract}

\date{\today}

\maketitle

Of all physical phenomena, light is perhaps the most crucial to our perception and understanding of the world. Its ability to transmit information regarding matter has long been harnessed as a central tool for the study of the natural world. Attempts to understand the nature of light stretch back to antiquity, with the majority of Hellenic philosophers espousing an `extramission' theory of vision, in which light propagated out from the eyes in order to `touch' objects \cite{Zubairy2016}. This theory was only disproven in the 11th century AD by Alhazen, who invented the first \emph{camera obscura} in the process \cite{AlKhalili2015}. In later years, Newton's determination to confirm his own theory of optics \cite{newton2012opticks, NewtonOptics1,NewtonOptics2} extended to self-experimentation, using a bodkin to deform the back of his eyeball \cite{darrigol2012a, bryson2003a}. A complete description of classical light was however not achieved until the work of Maxwell in the 1870s \cite{maxwell1954a}. 

In the modern era, as our understanding of light has progressed, it has assumed a role of central importance in both industry and research. In this context, light is not only a subject of fundamental study, but a tool for the manipulation of matter \cite{Lidar2004, PhysRevLett.118.083201,gross1993inverse,Chen1995,Chen1997,PhysRevA.98.043429,mccaul2020driven,mccaul2020controlling}, a medium of computation \cite{Pierangeli2021, Shen2017,Opala2019,Hamerly2019, Pappu2002, Bogaerts2020}, and a probe of physical systems. It is often assumed that the ultimate limitations of light for imaging and other optical experimentation correspond to the highest frequency in its Fourier decomposition (e.g. the diffraction limit). In some cases however, it is possible to construct a \emph {superoscillation} (SO) - a band-limited signal which has a temporal or spatial region in which it oscillates at a frequency faster than its largest Fourier component \cite{Berry_2019}.

The term superoscillation was first coined by Michael Berry \cite{BerryFourier,Rogers2020}, in reference to work by Aharonov, Bergman, and Lebowitz \cite{MR163614}.  Today superoscillations constitute a rapidly developing field of study in both mathematics \cite{MR2826550, MR3079857} and physics \cite{MR1082112, Zheludev2021}. The main hinderance to the application of superoscillations has been their minute intensity as compared to full pulse, but advances in the general sensitivity of measurements means that the relative amplitudes of superoscillations are no longer an impediment to their study and application \cite{Berry_2019}.  These applications cover a broad range of topics, including optical metrology \cite{optical_ruler}, super-resolution \cite{MR2233265, Lindberg_2012, Berry_2006}, super-transmission  \cite{Eliezer:14, Eberly_2016,zarkovsky_transmission_2020} and free-space plasmonics \cite{Yuan2019}. 

In the spatial domain, there have been a number of experimental observations of superoscillations \cite{Huang2009,Rogers2020, Rogers2012,Santiago2021}, which extend even into the THz optical frequency range and allow for imaging below the diffraction limit \cite{Pu}. This raises the tantalising possibility that nanoscale imaging with visible light may be made possible by superoscillation \cite{photonics5040056}. In the temporal domain THz acoustic superoscillations have been observed in superlattices \cite{brehm_temporal_2020}, while the envelopes of laser pulses have been manipulated to display superoscillatory behaviour (relative to the envelope frequency) \cite{PhysRevLett.119.043903, Eliezer:18}.  Up to this point however, there has been no direct observation of a time domain THz optical superoscillation, despite the availability of THz radiation which can be both characterised and manipulated in the time domain \cite{time_domain_spectroscopy}. Much like in the spatial domain, temporal superoscillations have tremendous potential to broaden the domain of useful operation for current technology. One could for example use superoscillations to perform \emph{superspectroscopy} at higher frequencies, enhancing the range of spectroscopically accessible frequencies by currently available light sources.

At present, the prescriptions for directly constructing superoscillations (as opposed to achieving them via the design of lenses \cite{Yuan2017} or filters \cite{Lindberg_2012}) have been based on the analysis of band limited functions, but transitioning to experimentally viable approximations of such functions is fraught with difficulties \cite{Berry2016,Tang2016,Katzav2016}. Despite the sensitivity of purposely constructed superoscillations, there has been a preponderance of `accidental' superoscillations  observed both in random functions \cite{BerryFourier} and more structured fields \cite{Kozawa:18,Stafeev:16}. One might therefore ask if there exists a simple heuristic for generating superoscillatory fields through the superposition of experimentally available pulses, which avoids the difficulties of explicitly constructive techniques. Here we present just such a method for the generation of time domain superoscillations, and apply it to achieve a first realisation of an optical time-domain superoscillation in the THz regime.

\subsection*{Superoscillations as destructive interference \label{sec:superoscillations}}
\begin{figure}
    \centering
    \includegraphics[width=0.5\textwidth]{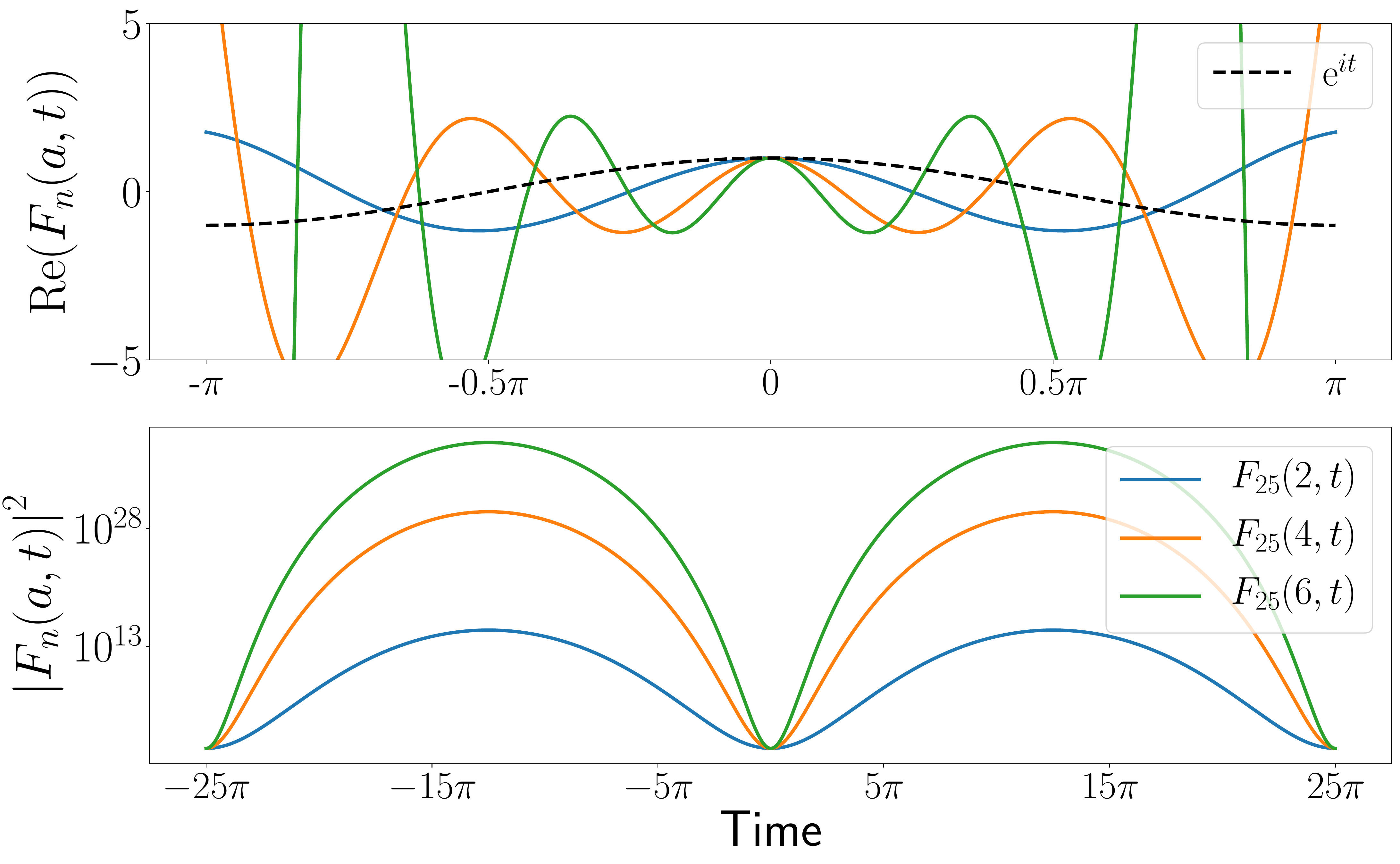}
    \caption{ Comparison of superoscillating functions (top panel) to their intensities (bottom panel). Comparison between the two panels readily demonstrates that the region where the magnitude $\left|F_n(a,t)\right|^2$ is minimised corresponds to a superoscillatory window where the function oscillates faster than its greatest Fourier component $k_{0}=1$ (shown as dashed line). The precise frequency achieved by the superoscillation is a function of the parameter $a$.
     }
    \label{fig:superexample}
\end{figure}

The phenomenon of superoscillations has typically been studied in the context of the function \cite{BerryFourier,Berry_2006}
\begin{equation}
\label{eq:mainfunc}
    F_n(a,t)=\left[\cos\frac{t}{n}+ia\sin\frac{t}{n}\right]^n.
\end{equation}
The behaviour of this function for small $t$ (or equivalently large $n$) can be obtained by Taylor expanding to first order:
\begin{equation}
\label{eq:Fapprox}
    F_n(t,a) \approx \left(1+ia\frac{t}{n}\right)^n \approx {\rm e}^{iat}. 
\end{equation}
When $a=1$ this expression is exact for any $n$, but in the case of $a>1$, it produces the counterintuitive behaviour that $F_n(a,t)$ oscillates faster than its highest frequency Fourier component -- $e^{it}$. To show this, let us re-express Eq.~\eqref{eq:mainfunc} as a Fourier series. The most direct route to achieving this is to express the trigonometric functions as exponentials before performing a binomial expansion. This leads directly to the Fourier expansion of $F_n(a,t)$:
\begin{align}
    \label{eq:FFourier}F_n(a,t)&=\sum_{j=0}^n A_j{\rm e}^{ik_j t}, \\
    \label{eq:prefactor} A_j&=\frac{1}{2^n}{n \choose j} (1+a)^{n-j}(1-a)^j, \\
    k_j & = 1-2 \frac{j}{n}.
\end{align}
Regardless of the value of $a$ chosen, the highest frequency in the Fourier decomposition will be $k_0=1$. Hence, for $a>1$, the function close to the origin will ``superoscillate'' at a frequency greater than its largest Fourier component. This is illustrated in Fig. \ref{fig:superexample} (top panel), where the frequency of the superoscillation relative to its largest Fourier component increases with $a$. 


A function of the type described by Eq.\eqref{eq:mainfunc} is far from the only way to obtain a superoscillation however. Careful examination of the secondary properties $F_n(a,t)$ can be used to formulate an alternative prescription for generating super oscillations. Here we focus on the fact that superoscillations occur in the region where $\left|F_n(a,t)\right|^2$ is minimised, as shown by Fig. \ref{fig:superexample} (bottom panel). This phenomenon can be understood heuristically by noting that the magnitude of the superoscillation will be of order 1 by Eq.\eqref{eq:Fapprox}. Examination of the Fourier components via Eq.\eqref{eq:prefactor} however shows individual frequencies may have much larger amplitudes.  The size of these components can only be reconciled with the superoscillation if there is near total cancellation of the individual field components around $t=0$. This means that \emph{superoscillations are a product of destructive interference}. This fact allows us to use the proxy of destructive interference as a figure of merit for the construction of superoscillations.

\section*{Results}

Let us now apply this logic to the problem of generating a superoscillating  optical field $E(t)$ using a set of experimentally accessible, time-limited  $E_j(t)$ waveforms with different central frequencies $\omega_j$. Each waveform has an amplitude $a_j$ and time delay $\tau_j$ such that the total field is given via 
\begin{equation}\label{EqSuperPositionField}
    E(t) = \sum_{j=1}^N a_j E_j(t-\tau_j).    
\end{equation}
We wish to select our parameters such that the intensity of this waveform is minimised over some time window  $t\in [-T_{\rm SO},T_{\rm SO}]$ which we shall refer to as the superoscillatory region. This immediately presents two difficulties if one varies amplitudes. First, in order to avoid the trivial solution $a_j=0$, it is necessary to condition any minimisation to avoid this outcome. Moreover, amplitude modulation is guaranteed to reduce the integrated intensity of the full waveform, making experimental detection more challenging.

\begin{figure}
    
    \centering
    \includegraphics[width=0.5\textwidth]{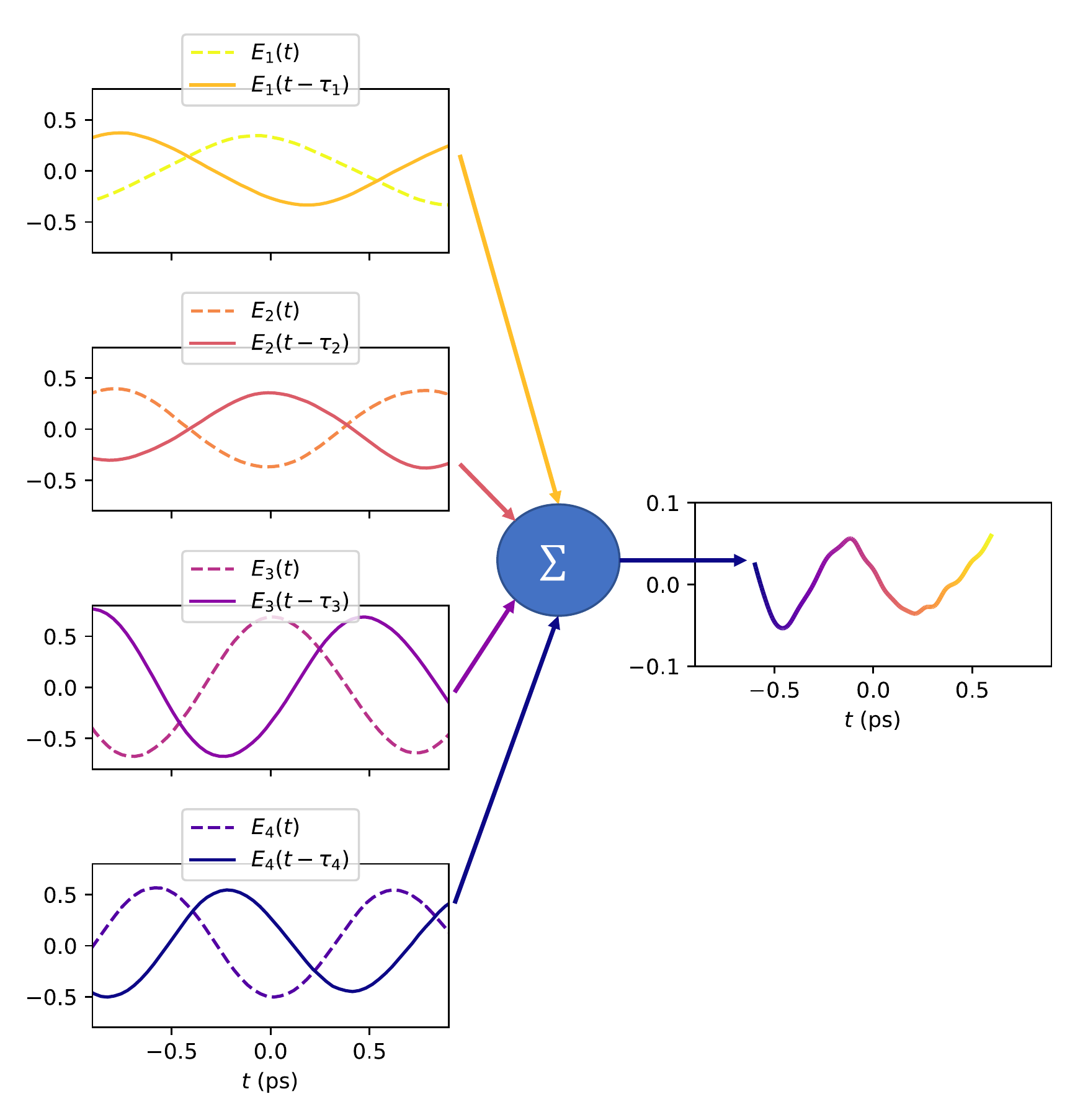}
    \caption{\label{fig:example_combo} In order to generate superoscillations, each component waveforms is shifted in time by an amount which minimises Eq.\eqref{eq:costfunction}. The combination of these shifted waveforms then creates a superoscillation within the desired region.}
\end{figure}

For this reason, we consider only variations of the time delays $\tau_j$ when minimising intensity within the superoscillatory region (i.e. setting $a_j=1$). Explicitly, we wish to minimise the objective function
\begin{align}
\label{eq:costfunction}
    I(\{\tau_j\}) = \int_{-T_{\rm SO}}^{T_{\rm SO}} \Bigg[ \sum_{j=1}^N E_j(t - \tau_j) \Bigg]^2 {\rm d}t 
\end{align}
by varying the time delays $\tau_j$. Given the different central frequencies of the component waveforms, the integrated intensity of the combined beam will remain roughly constant as time delays are varied, but by minimising that intensity in some temporal region, we expect to generate superoscillatory behaviour. Fig.\ref{fig:example_combo} shows an example of this, where each component waveform is shifted in time such that a superoscillation is generated by their superposition. 

This objective function can be minimised in any number of ways, e.g. via gradient descent \cite{fletcher2000practical}. The cost function itself is non-convex, meaning a number of local minima may be obtained depending on the initial guess for time delays. Finding the global minimum is not imperative however, as given the trade-off between the superoscillation amplitude and frequency, such a solution is likely to correspond to a superoscillation whose amplitude is beyond the limit of experimental detection. It is therefore useful to have a number of minima when seeking the best balance between the amplitude and frequency of the superoscillation.

\begin{figure}
    
    \centering
    \includegraphics[width=0.5\textwidth]{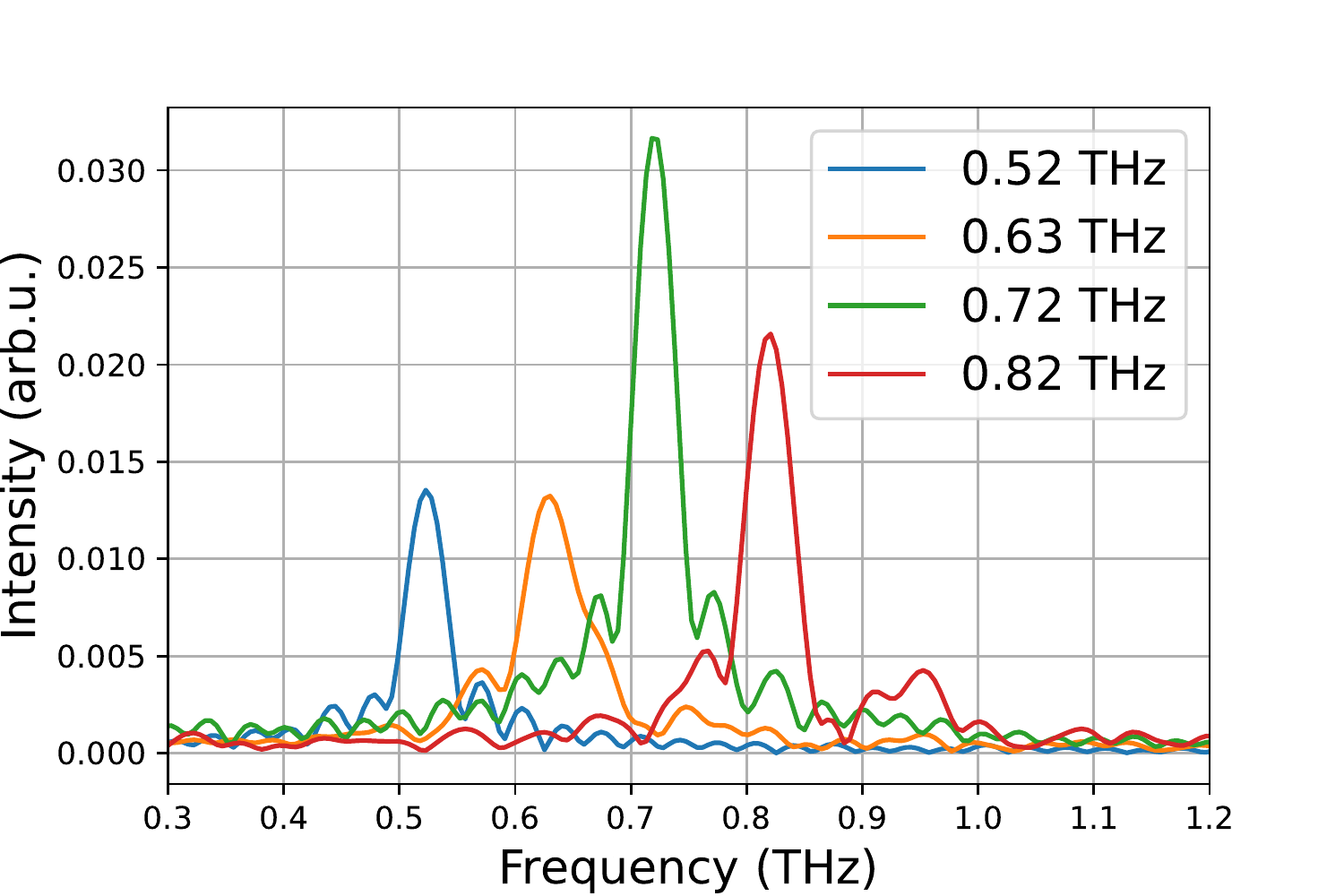}
    \caption{\label{fig:Fourier_components} Spectra of each of the four component waveforms used to generate the superoscillation. Each wave is labeled by its nominal frequency in THz.}
\end{figure}

As a first demonstration of the viability of this method for generating superoscillations, we combine four near-sinusoidal THz optical beams. Fig. \ref{fig:Fourier_components} shows the spectrum of each of these component waveforms. Note that while each measured waveform deviates from monochromaticity, the prescription for generating superoscillations is insensitive to the precise form of its constituent beams. Additionally, we set  $T_{\rm SO}=0.6$ ps, such that the superoscillatory region has a length corresponding to approximately one period of the highest frequency beam.

\begin{figure*}[t]
    \centering
    \includegraphics[width=0.9\textwidth]{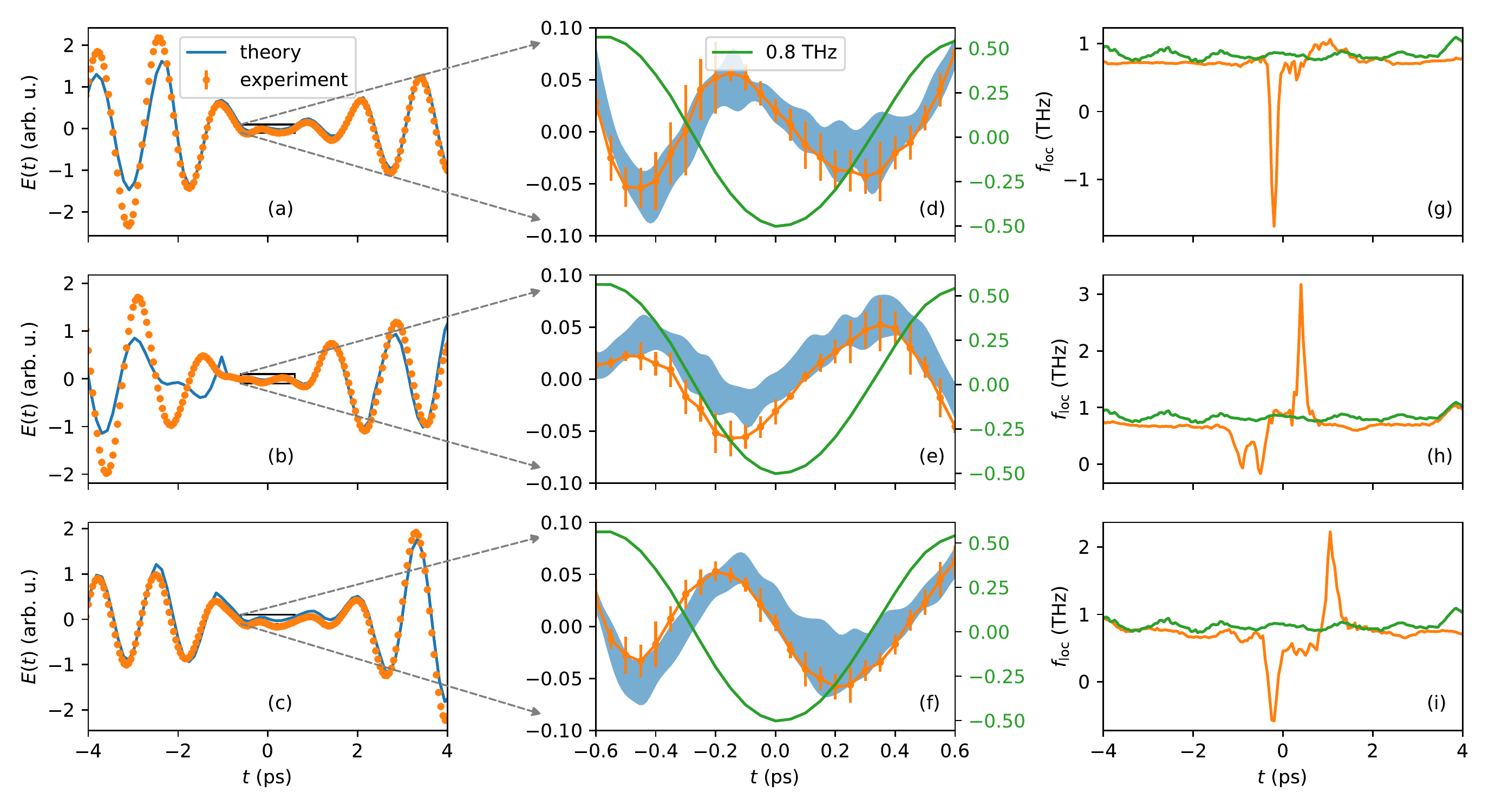}
    \caption{Three example superoscillations obtained via the minimisation of Eq. \eqref{eq:costfunction}. Left column (a, b, c) shows both the predicted and observed field over the full measurement window. The middle panels (d, e, f) highlight the superoscillatory region in which the superoscillation occurs, with comparison to the highest frequency component waveform.  The experimental data is in excellent agreement with the theoretical 95\% confidence limit (shaded blue area). The right column (g, h, i) plots the local frequencies of both the total waveform and its highest frequency component. Outside the superoscillatory region these are roughly equivalent, but within the superoscillatory region the combined waveform's local frequency increases sharply compared to that of the highest frequency component.}
    \label{fig:ObtainedSO1}
\end{figure*}

After measuring each component waveform $E_j(t)$ individually, a minimisation of Eq.\eqref{eq:costfunction} is performed. The time delays prescribed by this procedure are then implemented experimentally. Three example minima obtained from this process are shown in Fig. \ref{fig:ObtainedSO1}. As predicted, in each case superoscillations are clearly observed within the superoscillatory region [Figs.~\ref{fig:ObtainedSO1}(d,e,f)]. These superoscillations are not only of a sufficient amplitude for detection, but exhibit excellent agreement with the theoretical prediction used to generate them. 

The precise enhancement in frequency provided by the superoscillation may be quantified via a calculation of `local frequency'. This is defined in analogy with the local wavenumber used to characterise spatial superoscillations  \cite{Rogers_2013}, and is the gradient of the phase $f_{\rm loc}=\frac{{\rm d} \phi}{{\rm d}t}$.  This is obtained by Hilbert transforming the real valued signal $E(t)$ into an analytic form \cite{smith2007mathematics}, from which the field phase $\phi$ is extracted.  As Figs.~\ref{fig:ObtainedSO1}(g, h, i) show, for most of the duration of the combined waveform, its local frequency is approximately equal to its highest frequency component. Within the superoscillatory region however, the combined waveform's $f_{\rm loc}$ deviates sharply, experiencing a $\sim$ 2 fold increase relative to the highest frequency component. 
    %

\section*{Discussion and Outlook}
The ability to probe systems with optical signals underpins the ongoing development of quantum technologies, and superoscillatory signals have the potential to greatly expand the toolset available to researchers. The present work addresses the question of how to practically generate such superoscillations in the time domain. While the results presented here employ a specific methodology for producing superoscillations, the ultimate goal of the method - to minimise pulse intensity over a given time window - can be achieved with a great variety of techniques. This freedom means that the prescription for generating superoscillations may be tailored to enforce some desired secondary properties, depending on the application and specific technique employed. Indeed, perhaps the principal future challenge will be to develop methodologies that maximise the control available in the generation of superoscillations. Nevertheless, as has been shown here, even a relatively simplistic prescription is capable of producing clearly detectable optical superoscillations in the time domain. 

The potential applications of superoscillations cover a broad range of topics. Most obviously they extend the range of frequencies accessible by a given light source. This is of particular interest as researchers begin to explore dynamics in the attosecond regime \cite{RevModPhys.81.163}. At present the high frequency light necessary to probe this timescale is generated via highly non-linear effects such as high harmonic generation \cite{Ghimire2011,Ghimire2012,Silva2018}, but superoscillations offer an alternative platform for realising these light sources, without recourse to non-linear effects. 

Furthermore, recent developments in quantum control have demonstrated the highly malleable nature of driven systems, to the extent that one can force one material to `mimic' the optical response of another \citep{Campos2017, ball, optical_indistinguishability,mixing_mccaul}. Such manipulations require complex, broadband fields however, and superoscillations may facilitate the realisation of these control fields. 

Finally, the super-transmissive property of superoscillations \cite{Eliezer:14} means that they can be used to transmit light through media at frequencies that would ordinarily be absorbed. This has the potential to allow for the probing of materials in a range at which they would ordinarily be optically opaque. Taken together, these applications suggest that superoscillations - and the ability to generate them - have the opportunity to become vital tools in the armory of optical physics. 

\section*{Data Availability}
All experimental data, together with the minimisation procedure required to construct the superoscillations shown in Fig.~\ref{fig:ObtainedSO1} may be found at \url{https://github.com/dibondar/superoscilations/blob/master/superoscillations from experimental data/get superoscilations from 03-11-2022 data.ipynb}

\acknowledgments
This was was supported by by the W. M. Keck Foundation. D.I.B. and G.M. are also supported by Army Research Office (ARO) (grant W911NF-19-1-0377; program manager Dr.~James Joseph). The views and conclusions contained in this document are those of the authors and should not be interpreted as representing the official policies, either expressed or implied, of ARO or the U.S. Government. The U.S. Government is authorized to reproduce and distribute reprints for Government purposes notwithstanding any copyright notation herein.

\subsection*{Author Contributions}
G.M., D.L. and D.I.B. performed the theoretical analysis, while P.P., M.O.M. and D.T. designed the experimental setup and performed measurements. D.T. and D.I.B. supervised and coordinated this project, while G.M. wrote the initial manuscript draft, with the assistance of all other authors. G.M. and P.P. contributed equally to this work. 
\bibliography{bibliography}

\end{document}